\documentclass[twocolumn,preprintnumbers,amsmath,superscriptaddress,amssymb,aps]{revtex4-1}

\usepackage{graphicx}
\usepackage{dcolumn}
\usepackage{bm}
\usepackage{amsmath}
\usepackage{amssymb}
\usepackage{graphicx}
\usepackage{indentfirst}
\usepackage{booktabs}
\usepackage{multirow}
\usepackage{colortbl}

\selectfont

\begin{document}
\title{Progress on two-dimensional ferrovalley materials}

\author{Ping Li}
\email{pli@xjtu.edu.cn}
\address{State Key Laboratory for Mechanical Behavior of Materials, Center for Spintronics and Quantum System, School of Materials Science and Engineering, Xi'an Jiaotong University, Xi'an, Shaanxi, 710049, China}
\address{State Key Laboratory for Surface Physics and Department of Physics, Fudan University, Shanghai, 200433, China}
\author{Bang Liu}
\address{State Key Laboratory for Mechanical Behavior of Materials, Center for Spintronics and Quantum System, School of Materials Science and Engineering, Xi'an Jiaotong University, Xi'an, Shaanxi, 710049, China}
\author{Wei-Xi Zhang}
\address{Department of Physics and Electronic Engineering, Tongren University, Tongren 554300, China}
\author{Zhi-Xin Guo}
\email{zxguo08@xjtu.edu.cn}
\address{State Key Laboratory for Mechanical Behavior of Materials, Center for Spintronics and Quantum System, School of Materials Science and Engineering, Xi'an Jiaotong University, Xi'an, Shaanxi, 710049, China}

\date{\today}

\begin{abstract}
The electron's charge and spin degrees of freedom are at the core of modern electronic devices. With the in-depth investigation of two-dimensional materials, another degree of freedom, valley, has also attracted tremendous research interest. The intrinsic spontaneous valley polarization in two-dimensional magnetic systems, ferrovalley material, provides convenience for detecting and modulating the valley. In this review, we first introduce the development of valleytronics. Then, the valley polarization forms by the $\emph{p}$, $\emph{d}$, and $\emph{f}$-orbit that are discussed. Following, we discuss the investigation progress of modulating the valley polarization of two-dimensional ferrovalley materials by multiple physical fields, such as electric, stacking mode, strain, and interface. Finally, we look forward to the future developments of valleytronics.
\end{abstract}

\maketitle

$\textbf{Keywords:}$ ferrovalley, valley polarization, two-dimensional materials, multi-field tunable

$\textbf{PACS:}$ 75.70.Tj, 75.85.+t, 75.30.Gw

\section{Introduction}
For a long time, finding new ways to manipulate electronics to develop new types of electronic components and the next generation of electronics has been a goal that people have been working on. The manipulation of electrons is often inseparable from the intrinsic freedom that it can have. For physics workers, the conceptual discovery and regulation of the unexploited degrees of freedom that electrons possess in solids is a frontier topic. As we all know, electrons have two intrinsic degrees of freedom, spin, and charge. Based on charge and spin degrees of freedom, people have exploited the extensive application of electronic technology and gradually mature spintronics. Do the electrons in the solid still have other degrees of freedom besides spin and charge? The answer is yes. In many crystal structures, such as the metal bismuth, the semiconductor silicon, germanium, gallium arsenide, and the insulating material diamond, there is a new degree of freedom available to the electron called the valley. \cite{1,2,3}

In the reciprocal space of Bloch electrons, the valley generally refers to the local energy extremum point in the valence band or conduction band. Although these valleys may degenerate in energy, they are not necessarily equivalent because these valleys tend to be entirely separated in momentum space.  For example, the six degenerate but not equivalent valleys (Dirac cones) in graphene are separated in the corner of the first Brillouin region labeled K or $\rm K'$. The transition of electrons between these discrete valleys needs the involvement of the assistance of impurities or phonons to satisfy the conservation of momentum, which significantly reduces the scattering probability of electronic states between valleys, so that the valley can be seen as the intrinsic discrete degrees of freedom of holes or electrons. The non-equilibrium distribution of electrons in these discrete valleys can be used to realize a new pattern of encoding information. For example, the two logical states of "0" and "1" can correspond to the different occupying states of electrons in these degenerate valleys. The intrinsic freedom of valley can be developed into a potential carrier for information storage and processing in future electronics, related field is called "valleytronics".

In 2004, two-dimensional graphene was experimentally prepared by mechanical stripping. \cite{4} Immediately afterward, the team of ethnic Chinese scientist Qian Niu proposed that the two Dirac cones opened a gap by breaking the spatial inversion symmetry of graphene. It will appear novel quantum transport behavior and valley circularly polarized light dichroism selective adsorption. \cite{5,6} The non-equivalent two valleys (K and $\rm K'$) have the opposite signs of Berry curvatures. Especially, the transportation coefficient and light selection caused by Berry curvature is the opposite, it provides the theoretical foundation for the light field modulation valley polarization. The theoretical scheme has not been experimentally verified, breaking the lattice symmetry of graphene has a huge challenge. Therefore, graphene is not an ideal valley material. Researchers have turned their attention to the 2H phase transition metal chalcogenides (TMDs) represented by MoS$_2$. \cite{7,8} These materials have hexagonal lattices similar to graphene and spatial inversion symmetry breaking, and the valley Hall effect and the optical selection rule of valley resolution can be observed experimentally. \cite{9,10}

In the 2H-TMDs system, although the unequal valley is separated in momentum space, its energy still remains degenerate and valley polarization cannot be realized. Similar to paraelectric and paramagnetic materials, such valley energy degenerate materials are called paravalley materials. The realization the degeneracy of the valley is very beneficial to the development valleytronics. Hence, researchers have gradually developed a series of methods to achieve valley polarization, such as magnetic atom doping, \cite{11,12,13,14} magnetic proximity effect,  \cite{15,16,17,18,19} external magnetic fields, \cite{20,21,22,23,24} and optical Stark effects. \cite{9,10,25,26} It is worth noting that the valley polarization is achieved by the above means. It is not an intrinsic property of the material, which is not conducive to regulation of the valley degree of freedom. In 2016, Duan et al. predicted the spontaneous valley polarization monolayer 2H-VSe$_2$ by first-principles calculation. \cite{27} They named the material ferrovalley material. Until now, a series of ferrovalley materials have been predicted by theory, such as LaBr$_2$, \cite{28} Nb$_3$I$_8$, \cite{29} FeX$_2$ (X = Cl, Br), \cite{30,31} VSi$_2$X$_4$ (X = N, P, As), \cite{32,33,34,35} XY (X = K, Rb, Cs; Y = N, P, As, Sb, Bi), \cite{36} Cr$_2$Se$_3$, \cite{37} MnPTe$_3$, \cite{38} C$_6$N$_3$H$_3$Au, \cite{39} GdI$_2$, \cite{40} BiFeO$_3$, \cite {41} CeI$_2$, \cite{42} and so on. Beyond that, the application of valley freedom cannot be separated from the modulation of the external field. The magnetic coupling of two-dimensional magnetic ferrovalley can effectively tune the reversal of valley polarization, but it is not conducive to the miniaturization and aggregation of valleytronics devices. Therefore, the modulation valley of electrical, stacking, strain, and interface is a new direction to develop valleytronics devices.

\section{Valley freedom and valley polarization}	
Two-dimensional materials with hexagonal lattice structures are essential carriers for the investigation of valleytronics. In the section, the basic concepts of valleytronics, valley freedom and valley polarization are introduced through the spatial inversion of broken graphene and 2H-TMDs materials.

\subsection{Valley freedom}
In  two-dimensional system, the concept of valley was first proposed in graphene. Graphene has honeycomb structure that contains two sets of hexagonal sublattices, conventionally known as the A, B sublattices, as shown in Fig. 1(a). \cite{43} The two sets of sublattices are connected by spatial inversion symmetry. The corresponding reciprocal lattice is also hexagonal, the orientation is equivalent to the positive lattice being rotated by 90$^{\circ}$. The first Brillouin zone is hexagonal, as illustrated in Fig. 1(b). In the hexagonal Brillouin zone, two distinct corners (K$_{\pm}$ = $\pm$4$\pi$/3ax) are connected by the time reversal symmetry. Other corners of the Brillouin zone are associated with K by translation of lattice vectors. The most striking feature of graphene comes from low energy excitation at K and $\rm K'$ point, where valence and conduction bands touch each other linearly, and these contact points are named Dirac point.

\begin{figure}[htb]
\begin{center}
\includegraphics[angle=0,width=0.7\linewidth]{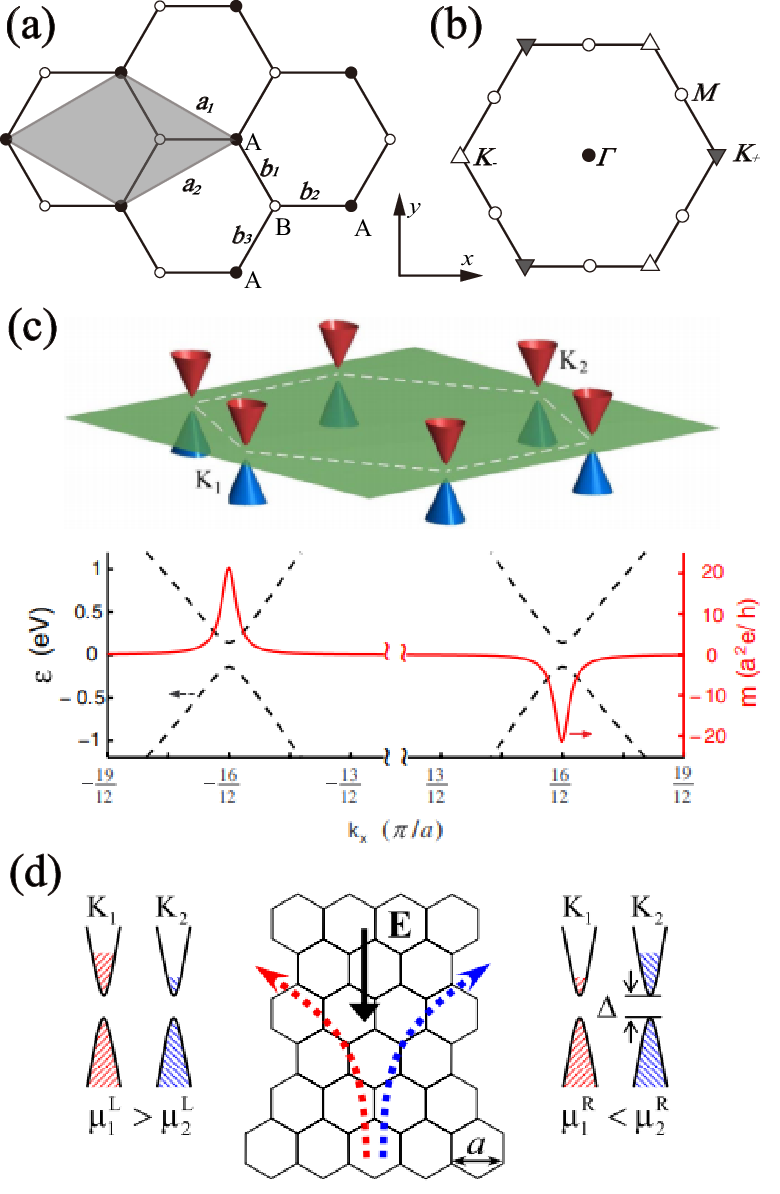}
\caption{(a) The honeycomb lattice of graphene, the sublattices A and B are represented by unfilled and filled circles, respectively. The a$_1$ and a$_2$ are lattice vectors. (b) Brillouin zone and high symmetry point. (c) The top panel is Band structures of graphene, the bottom panel is orbital magnetic moment of the conduction bands of graphene with broken spatial inversion symmetry. \cite{5} (d) Schematic diagram of valley Hall effect. The in-plane electric field generates transverse valley current, which results in a net valley polarization at the edge of the sample. \cite{5}}
\end{center}
\end{figure}

The semi-classical dynamical equation of Bloch electron under the external electric and magnetic fields can be expressed as \cite{44}
\begin{equation}
\dot{\textbf{r}} = \frac{1}{\hbar}\frac{\partial E_n(\textbf{k})}{\partial \textbf{k}} - \dot{\textbf{k}} \times \Omega_n(\textbf{k}),
\end{equation}
\begin{equation}
\hbar \dot{\textbf{k}} = -e \textbf{E} -e \dot{\textbf{r}} \times \textbf{B}
\end{equation}
where $\Omega_n(\textbf{k})$ is used to describe the Berry curvature of Bloch electron. For the system with spatial inversion symmetry, the Berry curvature meets $\Omega_n(\textbf{k}) = \Omega_n(\textbf{-k})$; for the system with time reversal symmetry, the Berry curvature meets $\Omega_n(\textbf{k}) = -\Omega_n(\textbf{-k})$; for the systems protected by both spatial and time reversal symmetry, the Berry curvature is zero everywhere. \cite{45} The Berry curvature is equivalent to the pseudo-magnetic field. When the non-zero Berry curvature exists in the system, the carrier motion will be affected, resulting in the valley Hall effect. Niu et al. found that the K$_1$ (K$_+$, K) and K$_2$ (K$_-$, $\rm K'$) valleys have opposite non-zero Berry curvature and orbital magnetic moments in graphene with the broken spatial inversion symmetry. \cite{5} As shown in Fig. 1(c), the energy extreme point will appear in the six corners of the first Brillouin zone, namely the K$_1$ and K$_2$ valleys. The K$_1$ and K$_2$ valleys are correlated by time reversal symmetry. As shown in Fig. 1(d), the carriers of K$_1$ and K$_2$ valleys have the same transverse velocity and opposite direction under the applied electric field. They turn around on both sides, respectively. That's the valley Hall effect. However, graphene is protected by both spatial inversion symmetry and time reversal symmetry. Therefore, it isn't easy to distinguish different valleys and manipulate valley degrees of freedom experimentally, which is greatly limited in practical applications.

\subsection{Valley polarization}
Due to the difficultly distinguishing the different valleys and manipulating valley degrees of freedom in graphene, 2H-TMDs materials are regarded as more ideal platform for valleytronics investigation. As shown in Fig. 2(a, b), unlike graphene, which only consists of a single planar carbon atom. Monolayer TMDs unit cell are triangular prism composed of two different atoms of transition metal and sulphur, which have naturally broken spatial inversion symmetry. \cite{7} But similar to graphene, monolayer TMDs also has a hexagonal honeycomb structure. Therefore, the valley of degenerate but not equivalent form at the K and $\rm K'$ points due to broken spatial inversion symmetry.

\begin{figure}[htb]
\begin{center}
\includegraphics[angle=0,width=1.0\linewidth]{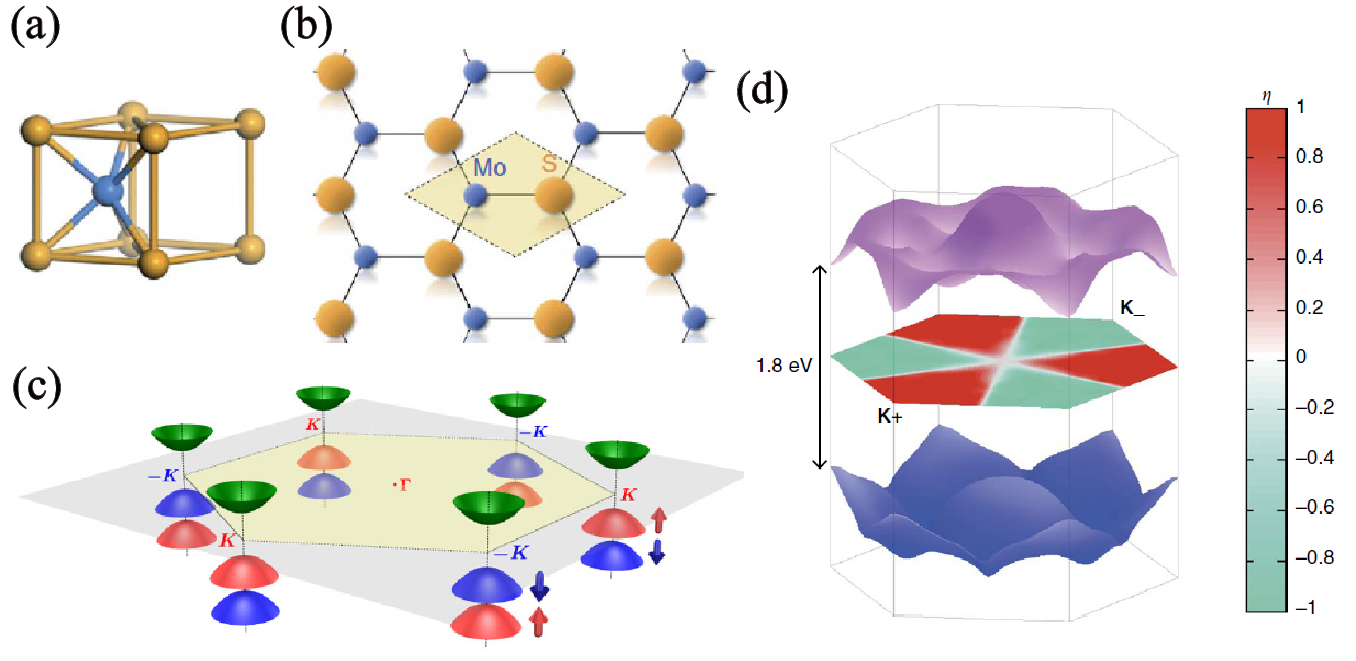}
\caption{(a) Coordination environment of Mo (blue ball) in the structure. The golden balls depicted S atom. \cite{7} (b) Top view of the MoS$_2$ monolayer lattice. \cite{7} (c) Valley distribution diagram at the K points. \cite{8} (d) Top valence band is shown blue, and bottom conduction band is shown pink. The Brillouin zone represent the center hexagon, the degree of circular polarization is shown by the center hexagon, $\eta(\textbf{k})$. \cite{7}}
\end{center}
\end{figure}

In 2H TMDs system, although the unequal valley is separated in momentum space, its energy still remains degenerate and valley polarization cannot be realized. Recently, the band splitting is caused by the spin-orbit coupling (SOC) effect of $\emph{d}$ electron of heavy metal atom, as shown in Fig. 2(c). \cite{8} Cao et al. calculated that the split of the valence band maximum was 0.15 eV. \cite{7,46,47} Moreover, the spin splitting is reversed at K and $\rm K'$ point due to the time reversal symmetry protection. The valley separated in both momentum and energy space are known as valley polarization. It provides possible to achieve dynamic valley polarization by selective circular dichroism light. Valley polarization can be characterized by the polarization component of photoluminescence (PL) intensity:
\begin{equation}
\eta = \frac{PL(\sigma_+) - PL(\sigma_-)}{PL(\sigma_+) + PL(\sigma_-)}
\end{equation}
As shown in Fig. 2(d), the optical selectivity is absolute at K$_{\pm}$ points, $\eta (K_{\pm})$ = $\pm$1. More surprisingly, the selectivity is close to perfect across the entire valley region, only rapidly changing signs at the valley boundary. It means that the entire K$_+$ valley almost completely rejects $\sigma ^+$ photons, whereas the entire K$_-$ valley almost completely repels $\sigma ^-$ photons. It indicates that the adopted $\sigma ^+$($\sigma ^-$) will produce satisfactory valley polarization. The experimental results show that the valley polarization can't reach 100 $\%$ due to the participation of phonons. \cite{9} In addition, the dynamic valley polarization of optical transition usually disappears within a few picoseconds, which greatly limits the practical application of optical induced valley polarization in valleytronics devices. \cite{48}

\section{Intrinsic ferrovalley materials}
Although the degradation of the valley can be achieved by various external means in TMDs materials, the valley polarization cannot be maintained when the external field is removed. Therefore, it is imperative to search for materials with spontaneous valley polarization. In this section, we introduce the investigation progress of ferrovalley materials from forming valley polarization of the orbit.

\subsection{$\emph{p}$-orbital ferrovalley materials}
Firstly, we introduce the $\emph{p}$-orbit ferromagnetic ferrovalley material. We predicted that the XY (X = K, Rb, Cs; Y = N, P, As, Sb, Bi) monolayer have intrinsic valley polarized quantum anomalous Hall effect. \cite{36} As shown in Fig. 3(a, b), it exhibits a honeycomb lattice with the point group D$_{3h}$ and space group P$\bar{6}$m2 (No. 187). The primitive unit cell is composed of an X and a Y atom, and its geometry is completely flat as that of graphene. The Y atom has unpaired electrons in its p orbital, leading to appear ferromagnetic. The intrinsic ferromagnetism breaks the time reversal symmetry of the system. As shown in Fig. 3(c, d), the combined interaction of ferromagnetism and SOC causes the monolayer XY to break the energetic degeneracy at the K and $\rm K'$ valleys, inducing spontaneous valley polarization.

\begin{figure}[htb]
\begin{center}
\includegraphics[angle=0,width=0.9\linewidth]{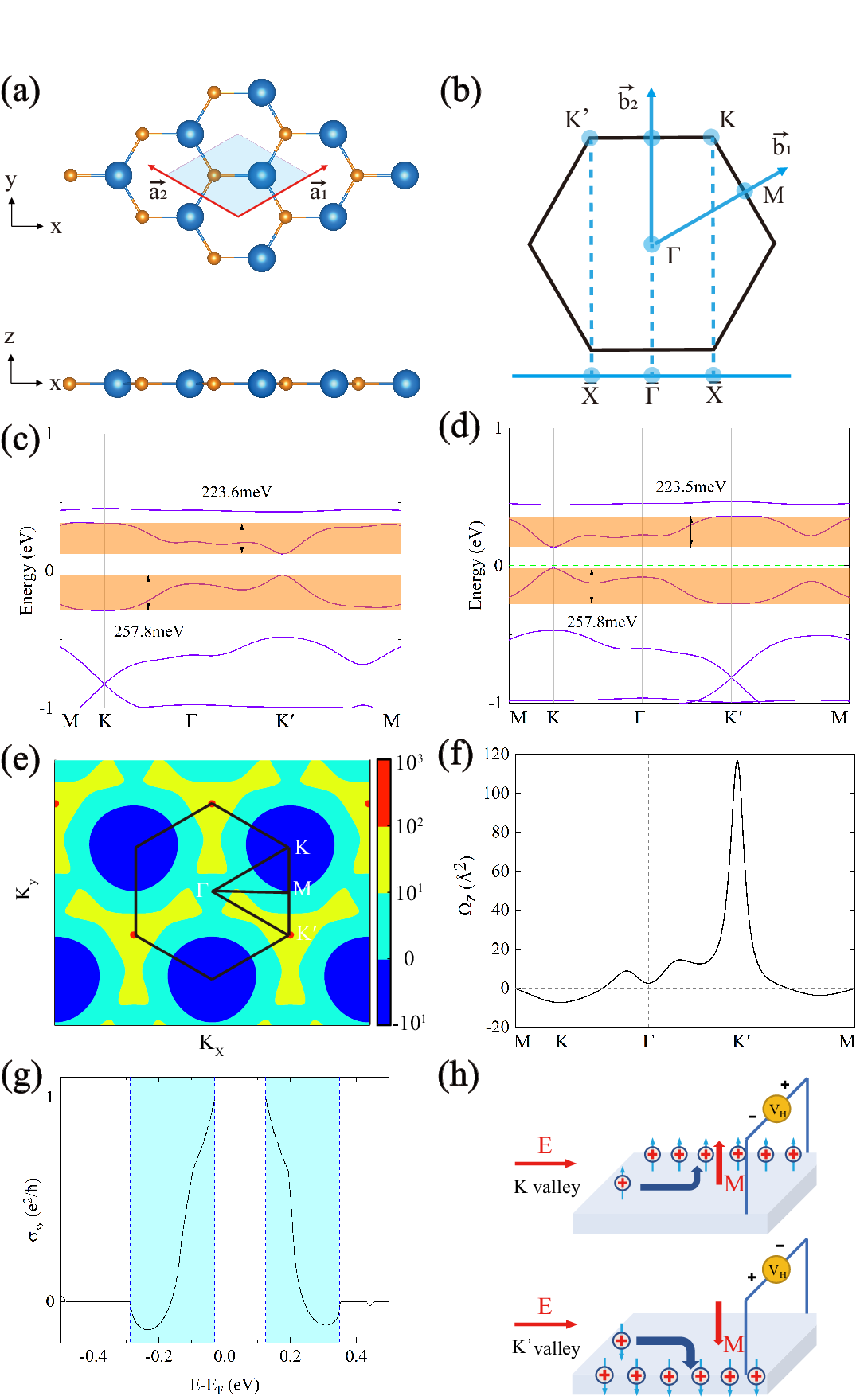}
\caption{(a) Top and side views of XY monolayer lattice structure. The X and Y atoms are represented by the blue and orange spheres, respectively. (b) The Brillouin zone of the honeycomb lattice. The band structure of CsSb including the SOC for the magnetic moment of Sb along the (c) z and (d) -z axis (out-of-plane), respectively. The Berry curvatures of CsSb over the two-dimensional Brillouin zone (e) and along the high symmetry line (f). (g) Calculated the AHC $\sigma_{xy}$ as a function of Fermi energy for CsSb monolayer. The two light blue shades indicate the valley split between K and $\rm K'$ valleys. (h) Schematic diagram of tunable valley polarized quantum anomalous Hall effect in hole doped CsSb monolayer at the K and $\rm K'$ valleys, respectively. The hole are represented by + symbol. Upward and downward arrows denoted the spin up and spin down carriers, respectively. \cite{36} }
\end{center}
\end{figure}

To understand the mechanism of the ferrovalley effect in XY, we take CsSb as an example. We try to build an effective Hamiltonian model based on the density functional theory (DFT) calculations. The conduction band minimum (CBM) and valence band maximum (VBM) are mainly consisted of Sb p$_x$ and p$_y$ orbitals including the SOC effect. Considering that the little group of K and $\rm K'$ points belongs to D$_{3h}$ in the out-of-plane magnetic configuration, we employed $|\psi_v ^\tau \rangle$ = $\frac{1}{\sqrt{2}}$$(|p_x\rangle + i\tau|p_y)\otimes |\downarrow \rangle$, $|\psi_c ^\tau \rangle$ = $\frac{1}{\sqrt{2}}$$(|p_y\rangle + i\tau|p_x)\otimes |\downarrow \rangle$ as the orbital basis for the CBM and VBM, where $\tau = \pm1$ shows the valley index referring to the K/$\rm K'$. Since the CBM and VBM belong the same spin channel, we take the SOC effect as the perturbation term, which is
\begin{equation}
\hat{H}_{SOC} = \lambda \hat{S}\hat{L} = \hat{H}^0 _{SOC} + \hat{H}^1 _{SOC}
\end{equation}
where $\hat{S}$ and $\hat{L}$ are spin and orbital angular operators, respectively. $\hat{H}_{SOC}^{0}$ and $\hat{H}_{SOC}^{1}$ denote the interaction between the same spin states and between opposite spin sates, respectively. For the CsSb monolayer, the single valley is consisted of only one spin channel, and the other spin channel is far from the valleys. Hence, the term $\hat{H}_{SOC}^{1}$ can be neglected. In addition,  $\hat{H}_{SOC}^{0}$ can be written in polar angles
\begin{equation}
\hat{H}_{SOC}^{0} = \lambda \hat{S}_{z'}(\hat{L}_zcos\theta + \frac{1}{2}\hat{L}_+e^{-i\phi}sin\theta + \frac{1}{2}\hat{L}_-e^{+i\phi}sin\theta),
\end{equation}
In the out-of-plane magnetic configuration, $\theta$ = $\phi$ = 0, then the $\hat{H}_{SOC}^{0}$ term can be simplified as
\begin{equation}
\hat{H}_{SOC}^{0} = \lambda \hat{S}_{z} \hat{L}_z,
\end{equation}
The energy levels of the valleys for the VBM and CBM can be expressed as E$_v$$^ \tau$ = $\langle$ $\psi$$_v$$^ \tau$ $|$ $\hat{H}$$_{SOC}^{0}$ $|$ $\psi$$_v$$^ \tau$ $\rangle$ and E$_c$$^ \tau$ = $\langle$ $\psi$$_c$$^ \tau$ $|$ $\hat{H}$$_{SOC}^{0}$ $|$ $\psi$$_c$$^ \tau$ $\rangle$, respectively. Then, the valley polarization in the valence and conduction bands can be represented as
\begin{equation}
E_{v}^{K} - E_{v}^{K'} = i \langle p_x | \hat{H}_{SOC}^{0} | p_y \rangle - i \langle p_y | \hat{H}_{SOC}^{0} | p_x \rangle \approx \lambda,
\end{equation}
\begin{equation}
E_{c}^{K} - E_{c}^{K'} = i \langle p_y | \hat{H}_{SOC}^{0} | p_x \rangle - i \langle p_x | \hat{H}_{SOC}^{0} | p_y \rangle \approx \lambda,
\end{equation}
where the $\hat{L}_z|p_x \rangle$ = i$\hbar$$|p_y \rangle$, $\hat{L}_z|p_y \rangle$ = -i$\hbar$$|p_x \rangle$. The results show that the valley degeneracy splitting of the valence and conduction band is consistent with our DFT calculation.

As shown in Fig. 3(e, f), the Berry curvatures of the K and $\rm K'$ points have opposite signs, indicating the typical valley polarization feature. By integrating the Berry curvature over the Brillouin zone, one can further calculate the anomalous Hall conductivity (AHC). As shown in Fig. 3(g), a valley polarized Hall conductivity clearly exists in CsSb. Particularly, when the Fermi level lies between the CBM or VBM of the K and $\rm K'$ valleys, the valley polarized Hall conductance ${\sigma}_{xy}$ can be obtained. This result confirms the presence of valley polarized quantum anomalous Hall effect (VQAHE) in the CsSb monolayer. Besides, in the hole doping condition, when the magnetic moment direction of CsSb is in +z direction, the spin-up holes from the $\rm K'$ valley will be generated and accumulated on one boundary of the sample under an in-plane electrical field [upper plane of Fig. 3(h)]. On the other hand, when the magnetic moment is in -z direction, the spin up holes from the K valley will be generated and accumulated on the opposite boundary of the sample under an in-plane electrical field [lower plane of Fig. 3(h)]. This characteristic indicates that monolayer CsSb is an ideal candidate for the high-performance valleytronic devices.

\begin{figure}[htb]
\begin{center}
\includegraphics[angle=0,width=0.9\linewidth]{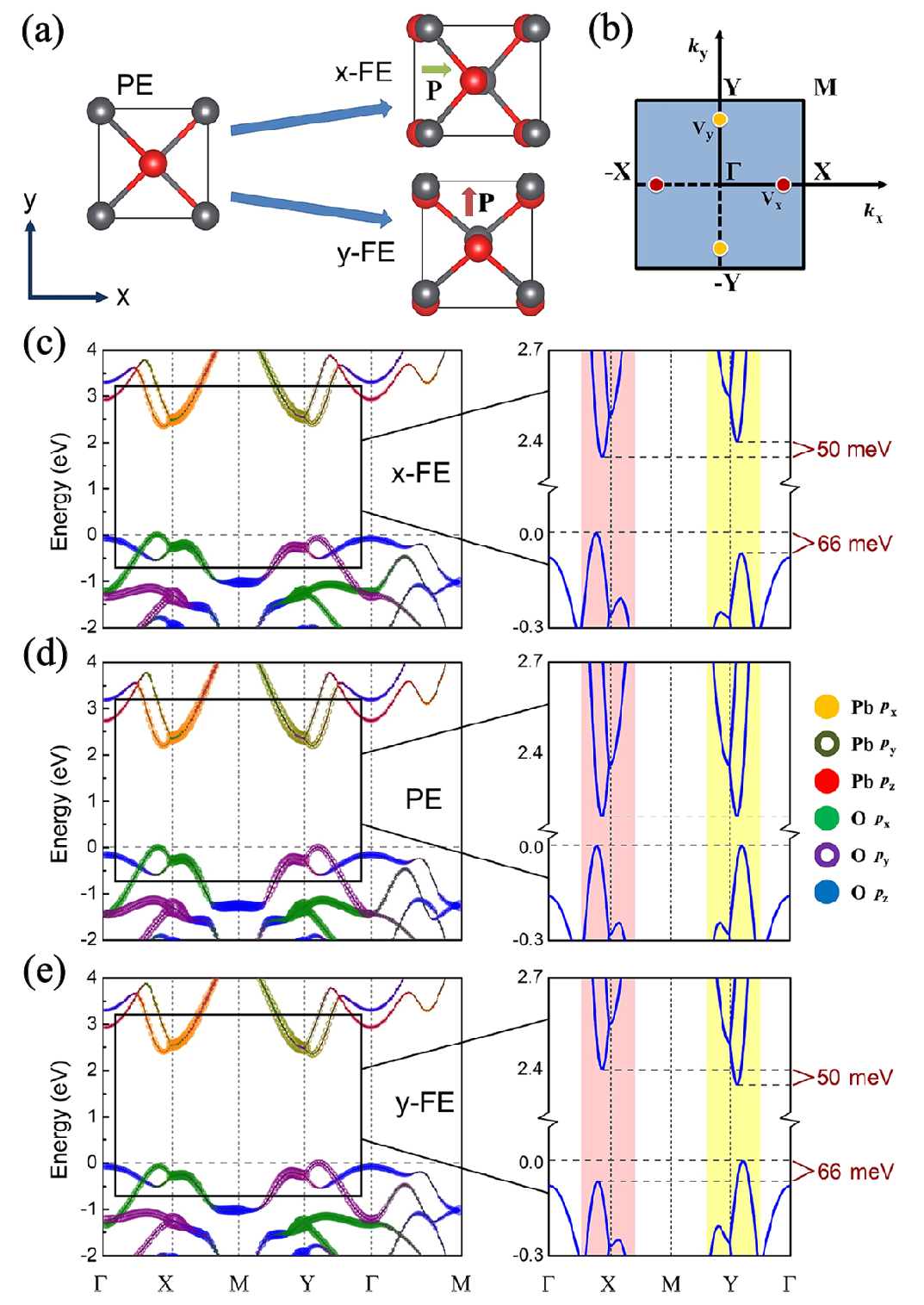}
\caption{(a) Schematic diagram of monolayer PbO in the non-polar PE phase and FE phase with polarization along the x direction (x-FE) and y direction (y-FE). (b) Brillouin zone of orthogonal lattice. Two valleys are signed V$_X$ and V$_Y$. (c-e) The band structures of monolayer PbO in x-FE, PE, and y-FE phases, respectively.  \cite{49} }
\end{center}
\end{figure}

In addition to the $\emph{p}$-orbital magnetically induced valley polarization, Jia et al. found that valley polarization can also be realized by the ferroelectric polarization of p-orbital (monolayer PbO) induced the broken spatial inversion symmetry. \cite{49} As shown in Fig. 4(a, b), the monolayer PbO is orthogonal lattice structure, different from the hexagonal lattice, the unequal valley of monolayer PbO is located on the $\Gamma$$\rightarrow$X and $\Gamma$$\rightarrow$Y high symmetric paths, respectively. When the system is the para-electric (PE) phase, the valley is energy degenerated. While the system switches to the ferroelectric (FE) phase, the intrinsic FE field breaks the spatial inversion symmetry induced valley degeneracy to disappear. As shown in Fig. 4(c-e), the VBM is mostly comes from  p$_x$ (around V$_X$) and p$_y$ (around V$_Y$) orbitals of O, while the CBM mainly contributed p$_x$ (around V$_X$) and p$_y$ (around V$_Y$) orbitals of Pb. For the two FE phases, spontaneous polarization actuates the structure into a ferrovalley state: the two pairs of valleys are no longer the same, but move away from the Fermi level in different ways. For the V$_X$ and V$_Y$ valley of PbO, the spontaneous valley polarization and linear polarization of light dependent optical selectivity. It not only indicates the potential of oxide structures for ultrathin electronic devices, but also provide the platform for future design and realization of two-dimensional multiferroics.

At present, ferrovalley materials of the $\emph{p}$-orbital are seldom studied. Usually, the valley splitting of the $\emph{p}$-orbital is very significant. Coincidentally, the room-temperature ferromagnetic $\emph{p}$-orbital CaCl monolayer has been experimentally synthesized on reduced graphene oxide membranes, \cite{CaCl} which confirms the possibility of synthesizing ferrovalley materials of the $\emph{p}$-orbital in experiments.

\subsection{$\emph{d}$-orbital ferrovalley materials}
The concept of ferrovalley was first proposed by Duan et al. in the d-orbital VSe$_2$ system. \cite{27} Its structure is the same as that of 2H-MoS$_2$, except for the presence of unpaired electrons in the 3d orbital of the V atom, which makes the monolayer 2H-VSe$_2$ ferromagnetic. Ferromagnetism and SOC effect together make monolayer 2H-VSe$_2$ decouple the energetically degenerated valleys at K$_{+}$ (K) and K$_{-}$ ($\rm K'$) points, and induce spontaneous valley polarization. The Hamiltonian of monolayer 2H-VSe$_2$ can be written as
\begin{equation}
H(\textbf{k}) = I_2 \otimes H_0(\textbf{k}) + H_{soc}(\textbf{k}) + H_{ex}(\textbf{k})
\end{equation}
where H$_{soc}$($\textbf{k}$), H$_{ex}$($\textbf{k}$) represent the SOC effect, and magnetic exchange interaction, respectively. The combination the SOC effect and ferromagnetic exchange interaction achieve spontaneous valley polarization without external field. The existence of spontaneous valley polarization makes the two unequal valleys of ferrovalley material that they have different optical band gaps. Optical calculation show that the system has valley-dependent rotation selectivity. As shown in Fig. 5(a), when the magnetic moment is along the z axis, the energetic degeneracy valley break. K$_{+}$ valleys can only be excited by left-circularly polarized light, while K$_{-}$ valley corresponds to right-circularly polarized light, and the absorption peak of right-circularly polarized light shows a little blue shift. As shown in Fig. 5(b), while the magnetic moment is along the -z axis, the optical band gap and the magnitude of valley polarization will be reversed accordingly. Fig. 5(c) is a diagram of the interband transition related to the optical band gap. Moreover, the two sets of valleys have opposite and unequal Berry curvatures for monolayer 2H-VSe$_2$, as shown in Fig. 5(d). This also makes ferrovalley materials have a measurable Hall voltage, which is similar to the anomalous Hall effect of ferromagnetic materials, called the anomalous valley Hall effect.  As shown in Fig. 5(e), based on the anomalous valley Hall effect, the magnetically writing and electrically reading memory devices are coming up. The binary information is stored by the valley polarization of the ferrovalley material, which can be modulated by the magnetic moment of the applied magnetic field.

\begin{figure}[htb]
\begin{center}
\includegraphics[angle=0,width=0.9\linewidth]{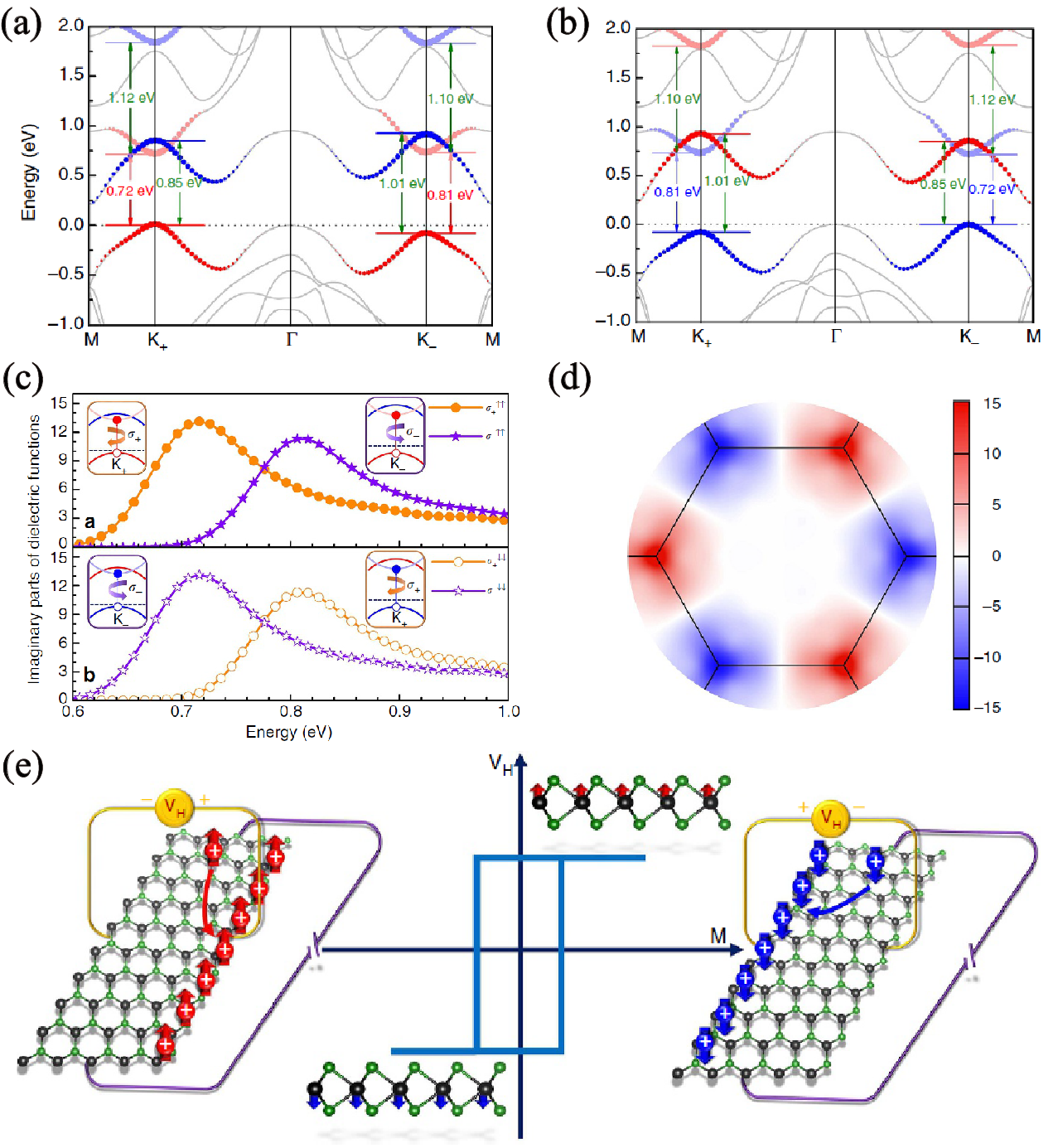}
\caption{The band structure with the SOC effect, the magnetic moments along the (a) z and (b) -z axis, respectively. (c) Schematic diagram of interband transitions dependent optical band gap. The imaginary parts of complex dielectric function $\varepsilon_2$ for monolayer VSe$_2$. $\sigma_{+}$ and $\sigma_{-}$ represent left and right handed radiation, respectively. (d) Contour maps of Berry curvature in the momentum space for bands is mainly occupied by d$_{xy}$ and d$_{x2-y2}$ orbits. (e) Diagram of data storage employing hole-doped ferrovalley materials based on anomalous valley Hall effect. \cite{27} }
\end{center}
\end{figure}

Moreover, we found that the Janus structure VSiXN$_4$ (X = C, Si, Ge, Sn, Pb) can realize multi-valley dependent topological phase transitions under the built-in electric field and strain. \cite{50} As shown in Fig. 6(a), with the atomic number of X increases, it shows that an interesting topological phase transition from valley semiconductor (VSC) to valley half-semimetal (VHSM), to valley polarized quantum anomalous Hall insulator (VQAHI), to the VHSM, to the VSC, and to valley metal (VM). Besides, it can be observed that the valence band valley of $\rm K'$ is obviously lower than K for the VSiCN$_4$, VSi$_2$N$_4$, and VSiGeN$_4$, while the conduction band valley of K$'$ is significantly higher than K for VSiSnN$_4$. Through the analysed orbital-resolved band structure, the VBM bands of VSiCN$_4$, VSi$_2$N$_4$, and VSiGeN$_4$ are dominated by V d$_{xy}$/d$_{x2-y2}$ orbitals, while the CBM bands are mainly contributed by d$_{z2}$ orbital of V atom. On the contrary, the VBM bands are mainly d$_{z2}$ orbital, while the CBM are primarily d$_{xy}$/d$_{x2-y2}$ orbitals for VSiSnN$_4$ and VSiPbN$_4$ monolayer. More importantly, it can be observed that the atomic number of the X atom induced band inversion between the d$_{z2}$ and d$_{xy}$/d$_{x2-y2}$ orbitals. Hence, the band structures with the vary strain and X atom is summarized in Fig. 6(b, c). In a word, the valley-related multiple topological phase transitions attribute to the change of the sign of the Berry curvatures at K and $\rm K'$ points. The band gap and topological property with the various strain for the VSiXN$_4$ monolayers are summarized in Fig. 6(d).

\begin{figure}[htb]
\begin{center}
\includegraphics[angle=0,width=0.9\linewidth]{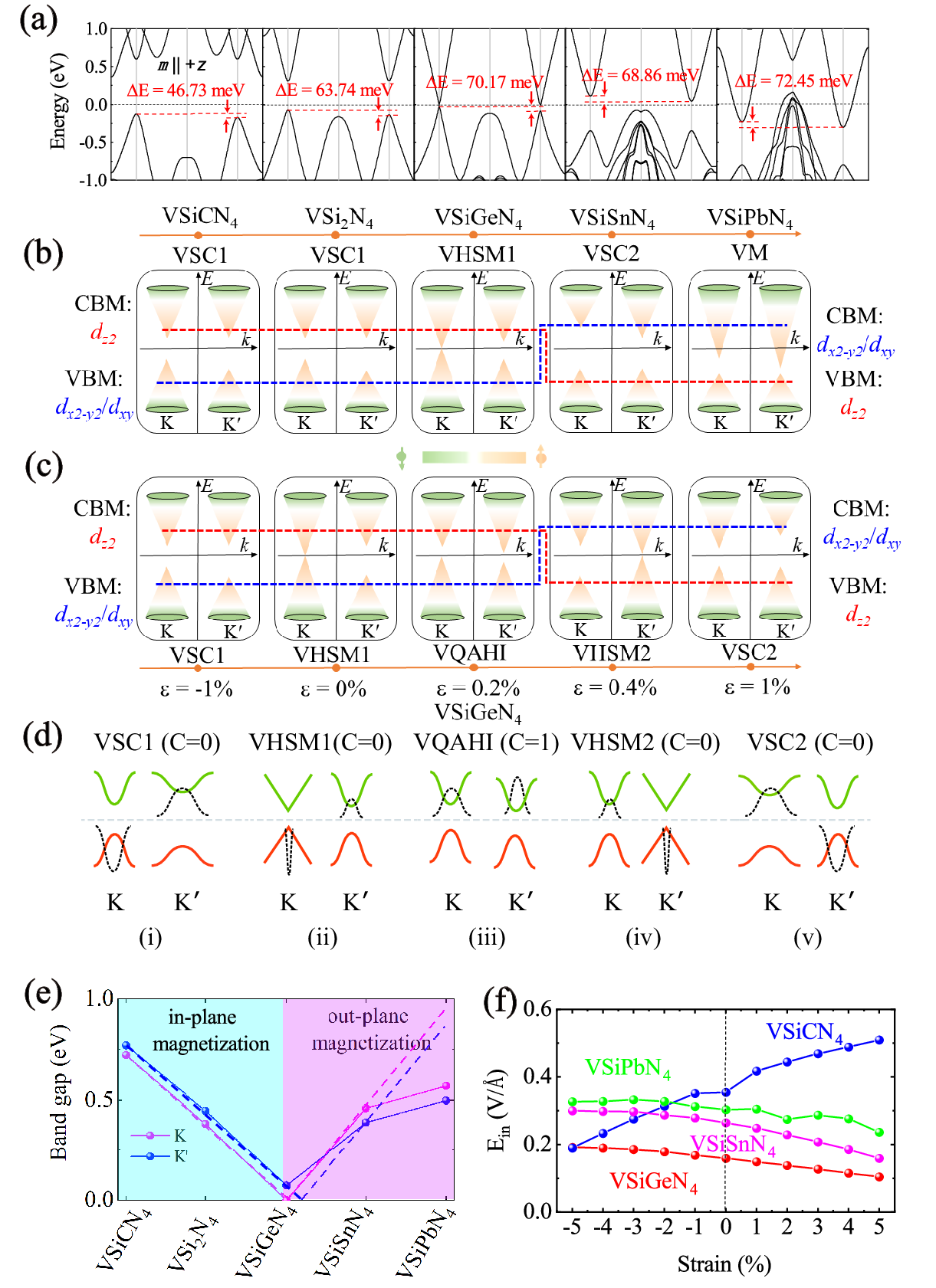}
\caption{(a) Band structure of the VSiXN$_4$ monolayers including the SOC effect. (b, c) Schematic diagrams of valley-dependent topological phase transitions, (b) With the atomic number of X element increasing, and (c) as a function of strain, respectively. (d) Evolution diagram of band structures and Berry curvatures of VSiXN$_4$ with different strains. The valence band, conduction band, and Berry curvatures of VSiXN$_4$ are showed the green solid line, red solid line, and dotted lines with different strains, respectively. (e) The band gap of VSiXN$_4$ at K and $\rm K'$ points. The dotted line is the fitting K and $\rm K'$ point gap variation trend. (f) The built-in electric field (E$_{in}$) as functions of the strain for VSiXN$_4$ monolayers. \cite{50} }
\end{center}
\end{figure}

Through a detailed analysis on the energy band variation with the atomic number of the X element, it is observed that the topological properties are intimately correlated to the gap at the K and $\rm K'$ points. The Fig. 6(e) shows the gap at the K and $\rm K'$ points with the atomic number of X increases, it found that the gap gradually decreases to zero in VSiGeN$_4$. The fitting curve displays that the band gap would reach zero when X is Ge, which indicates that this could be a phase transition point. It is well known that the built-in electric field was induced when atoms on the upper and lower surfaces are not equal. We define the built-in electric field as E$_{in}$ = ($\Phi_2$ - $\Phi_1$)/$\Delta h$, where $\Phi_1$ and $\Phi_2$ represent the electrostatic potential at the bottom and top of VSiXN$_4$, respectively. The $\Delta h$ represent the structural height of VSiXN$_4$. More surprisingly, the variation trend of the built-in electric field is entirely the same as that of the band gap of K and $\rm K'$, as shown in Fig. 6(f).It indicates that the intrinsic of topological phase transition is the strain and built-in electric field induces band inversion between the d$_{z2}$ and d$_{xy}$/d$_{x2-y2}$ orbitals at K and $\rm K'$ valleys. Our findings not only uncover the mechanism of multiple topological phase transitions, but also provides an new way for the multi-field manipulating the spin, valley, and topological physics. In recent years, the investigation of d-orbital ferrovalley materials has developed rapidly, a series of novel ferrovalley materials have been discovered, including MBr$_2$ (M = Ru, Os), \cite{51} VS$_2$, \cite{52} VSSe, \cite{53,54,55}, VClBr, \cite{56} RuCl$_2$, \cite{57} h-MNX (M = Ti, Zr, Hf; X = Cl, Br) and so on.

The $\emph{d}$-orbital ferrovalley materials have been extensively explored. These systems exhibit extraordinary properties, such as the valley spin Hall effect and valley polarized quantum anomalous Hall effect. The toughest problem is that the room-temperature magnetism cannot be realized in the experiment. Therefore, our unremitting efforts are still needed.

\subsection{$\emph{f}$-orbital ferrovalley materials}
Recently, Sheng et al. found the spontaneous valley polarization with $\emph{f}$ electrons in monolayer CeI$_2$. \cite{42} The monolayer CeI$_2$ exhibits a two-dimensional hexagonal lattice with the P$\bar{6}$m2 space group. The spatial inversion symmetry is broken in monolayer CeI$_2$. As shown in Fig. 7(a), the density of state uncovers that the magnetism originates from the Ce-5$\emph{d}$, Ce-4$\emph{f}$ and I-5$\emph{p}$ orbital electrons. Fig. 7(b, c) shows the band structure of monolayer CeI$_2$ including the SOC effect. The addition of the SOC effect eliminates the energy degeneracy of the K$_{+}$ and K$_{-}$ valleys in the top valence band, leading to an intriguing valley polarization. Moreover, it is found that the valley energy difference between the K$_{+}$ and K$_{-}$ valley depends the direction of magnetization. When the magnetic moment direction is along the +z axis, the valley polarization of spin up channel can reach up to 208 meV. Once the magnetization direction is reversed to the -z direction, the energy of the K$_{-}$ valley is higher than that of the K$_{+}$ valley, exhibiting a valley polarization of -208 meV in the spin down channel. In other words, the valley polarization and spin can be simultaneously flipped by reversing the magnetic moment of Ce atoms, which provides an efficient way to customize the valley characteristics of monolayer CeI$_2$ by modulating its direction of magnetization.

As shown in Fig. 7(d), it exhibits the calculated Berry curvature of monolayer CeI$_2$ as a contour map of the entire two-dimensional Brillouin zone. The K$_{-}$ and K$_{+}$ points have opposite signs and unequal magnitudes, meaning a robust valley-contrasting Berry curvature. When such p-type CeI$_2$ is magnetized along the +z direction, in the presence of an in-plane electric field, the positive Berry curvature drives the spin down holes from the K$_{+}$ valley to accumulate on the right side of the sample [see Fig. 7(e)]. Once the magnetization is reversed, the spin-up holes from the K$_{-}$ valley are transferred to the left side of the sample because of the action of negative Berry curvature. In addition to the anomalous valley Hall effect, it is also interesting to note from Fig. 7(e) that the charge and spin Hall effects exist in monolayer CeI$_2$, which is much easier to detected in experiments. Such excellent coexistence of charge, valley, and spin Hall current facilitates the integration of electronics, spintronics, and valleytronics. In addition to CeI$_2$, such LaX$_2$ (X= Br, I), \cite{28,59} GdXY (X, Y = Cl, Br, I; X $\neq$ Y), \cite{60} with $\emph{f}$-electron were reported. Due to the strong SOC of the $\emph{f}$-orbital, rare earth ferrovalley materials of the $\emph{f}$-orbital provide another possibility for the experimental studies.

\begin{figure}[htb]
\begin{center}
\includegraphics[angle=0,width=0.9\linewidth]{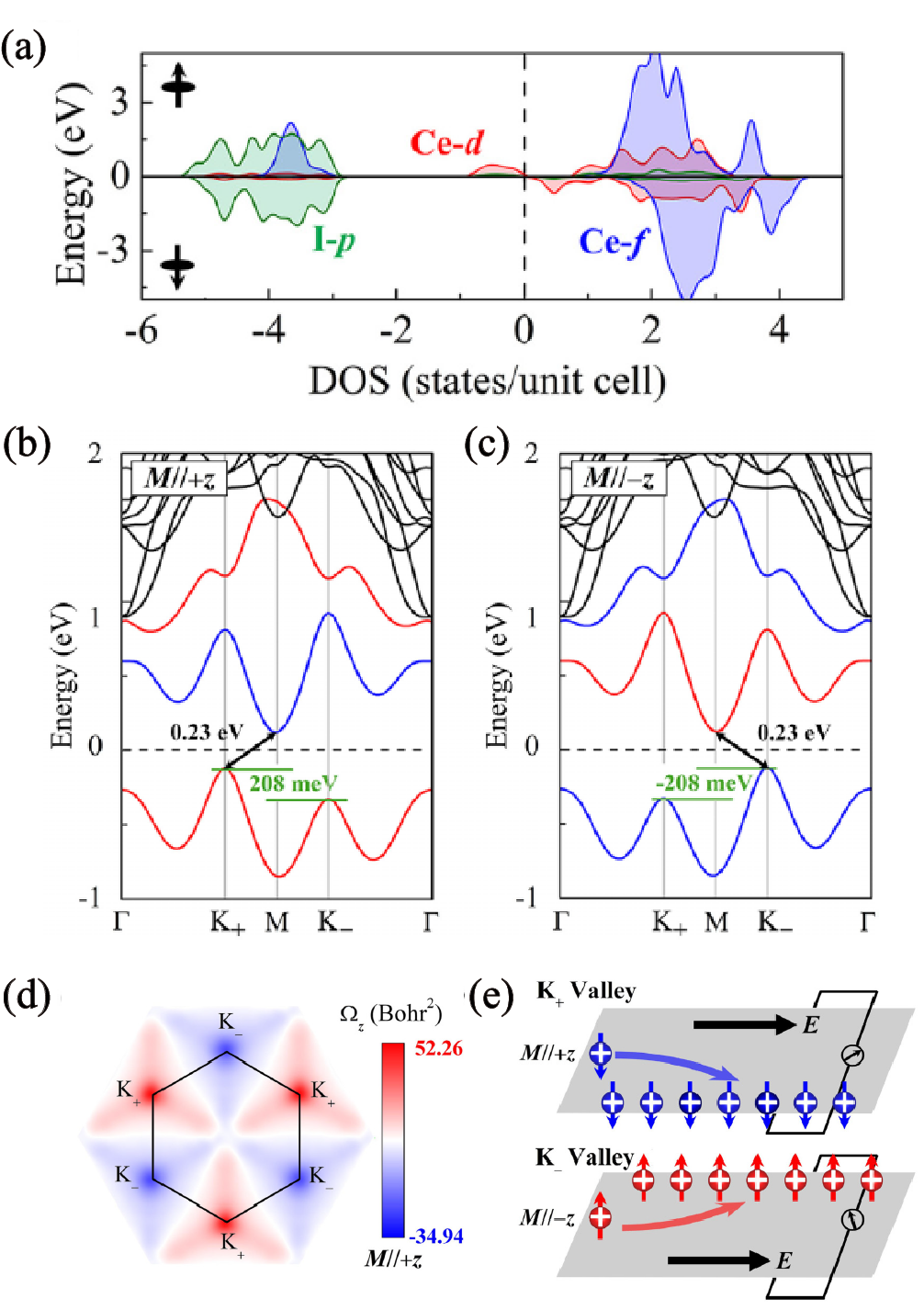}
\caption{(a) The orbital-resolved density of state for monolayer CeI$_2$. The SOC band structure with the (b) positive and (c) negative magnetic moment. (d) Berry curvature over the two-dimensional Brillouin zone. (e) Schematic of the anomalous valley Hall effect for the hole doped CeI$_2$ at the K$_{+}$ and K$_{-}$ valleys under an in-plane electric field. \cite{42}}
\end{center}
\end{figure}

\section{Multi-field tunable of two-dimensional ferrovalley materials}	
The two-dimensional ferrovalley materials maintain spontaneous valley polarization at the monolayer. The discovery of intrinsic ferrovalley materials has greatly promoted the development of valleytronics. The regulation of valley polarization of ferrovalley materials is the basis of information storage in valleytronic devices. The interlayer are bonded by a weak van der Waals interaction, which facilitates to tuning valley polarization by the multi-field (Electrical, interlayer stacking, strain, and interface). It can be realized in valleytronics and magnetic memory device applications.

\subsection{Electrical modulation}
The electrically controlled magnet has been widely investigated. \cite{61,62,Li1,Li2} Recently, the researchers found that the electricity can also modulate valley polarization well. \cite{63,64,65} Here, we mainly introduce two aspects: ferroelectric substrate and electric field.

\begin{figure}[htb]
\begin{center}
\includegraphics[angle=0,width=0.8\linewidth]{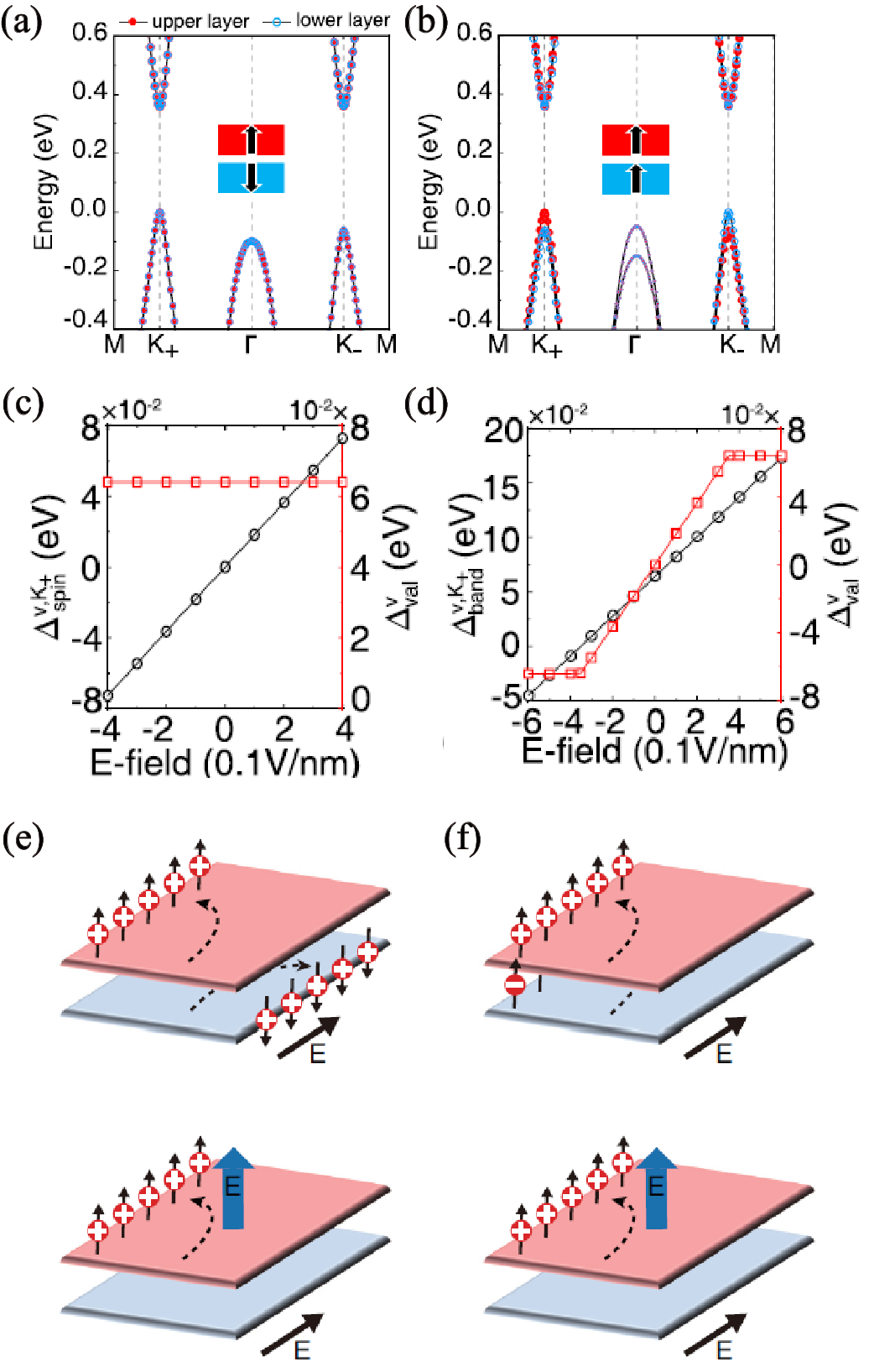}
\caption{(a) Band structures of the AA'-stacked VSi$_2$N$_4$ bilayers with the interlayer antiferromagnetic coupling. (b) Band structures of the AA'-stacked VSi$_2$N$_4$ bilayers with the interlayer ferromagnetic coupling. The red solid circles and blue hollow circles represent the atomic projected weights of electronic states on the V atoms from the upper and lower monolayer, respectively. (c, d) The band and valley splittings as functions of the electric field for the interlayer antiferromagnetic and antiferromagnetic coupling, respectively. The valley splitting is that of the valence bands, while the spin splitting corresponds to valence band at the K$_{+}$ valleys. (e, f) Schematic depictions of anomalous Hall currents in VSi$_2$N$_4$ bilayers with hole doping. (e) The interlayer antiferromagnetic coupling and (f) the ferromagnetic coupling. \cite{63} }
\end{center}
\end{figure}

Firstly, Liang et al. investigate that the electric field tune valley degeneracy splitting in VSi$_2$N$_4$ bilayer. \cite{63} They found that the valley splitting occurs in the interlayer antiferromagnetism, there are two degenerate valleys of the interlayer ferromagnetism at K$_{\pm}$ points, as shown in Fig. 8(a, b). For the interlayer antiferromagnetic coupling, the valley splittings does not change with the electric field, because the orbital components of the conduction/valence bands of the two valleys originate from the same monolayer, and the valley splitting of the monolayer is insensitive to the external electric field. Hence, the external electric field results in the spin splittings and has no effect on valley splittings. In addition, the sign of the spin splittings is decided by the field direction. When the field direction changes, the induced spin splittings are accordingly flipped (see Fig. 8(c)). On the contrast, as illustrated in Fig. 8(d), the size of the valley splitting lies on the electric field strength for the interlayer ferromagnetic coupling. While the positive electric field is in the range of 0-0.35 V/nm, the valley splitting is linearly increased, up to 64.1 meV. While the electric field continues enhancing, the valley splitting is unchanged. This trend change is due to the critical field of 0.35 V/nm that leads to a band exchange between the first and the second valence bands at the K$_{-}$ valley. Beyond the critical electric field, the first valence band of two valleys is contributed by the same layer. As a result, the valley-layer coupling is destroyed and the valley splitting is no longer changed with the electric field. In addition, due to the field-induced valley splittings, there is a transition from the direct band gap semiconductor to the indirect band gap semiconductor. Under the action of vertical electric field, the degeneracy splittings is highly adjustable for both the magnitude and the sign. These intriguing characteristics provide a practical way to designing energy-efficient devices based on coupling between multiple electronic degrees of freedom.

\begin{figure}[htb]
\begin{center}
\includegraphics[angle=0,width=1.0\linewidth]{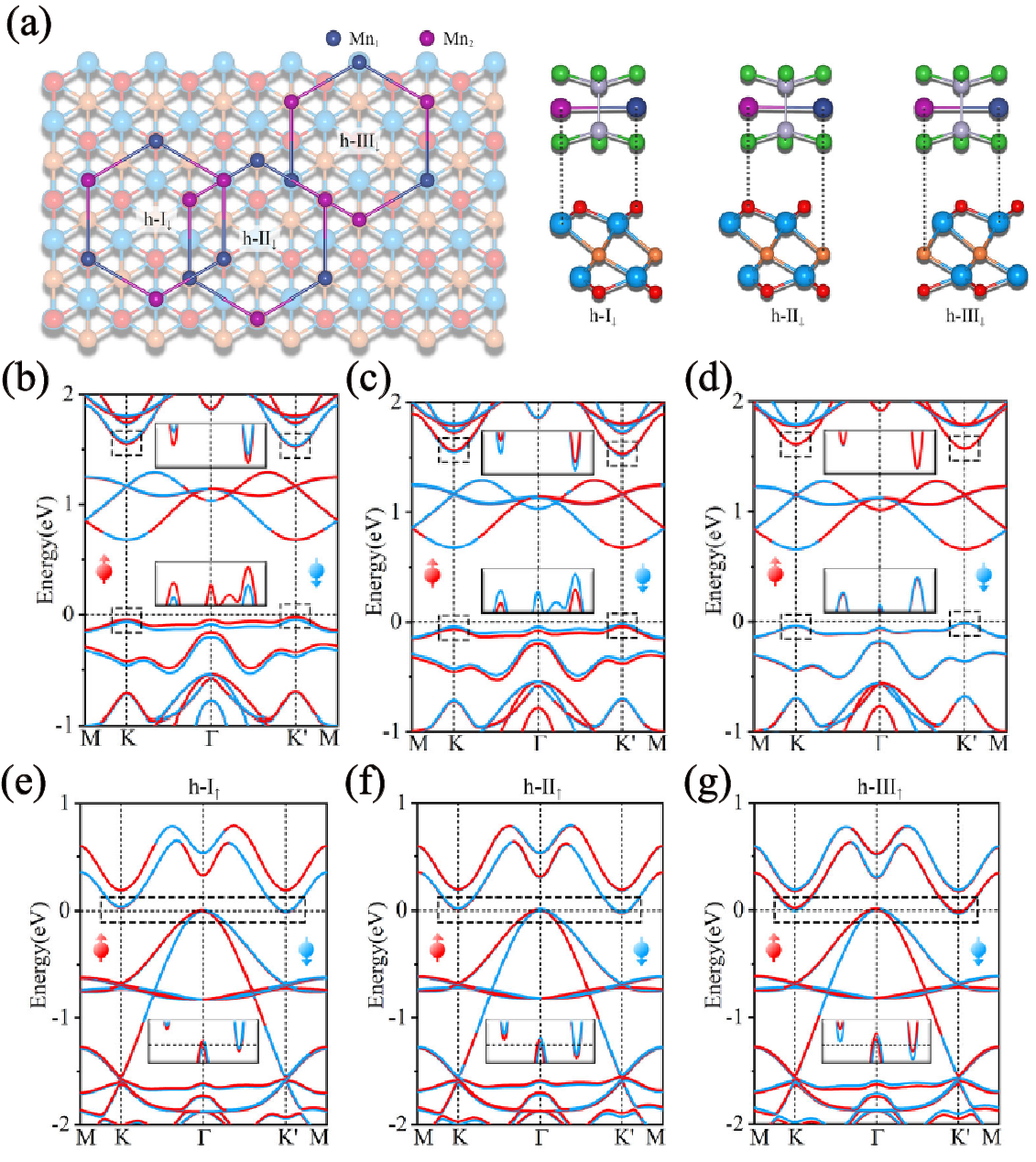}
\caption{(a) Crystal structures of MnPSe$_3$/Sc$_2$CO$_2$ h-I$_{\downarrow}$, h-II$_{\downarrow}$ and h-III$_{\downarrow}$. (b-d) Band structure with considering SOC of h-I$_{\downarrow}$, h-II$_{\downarrow}$ and h-III$_{\downarrow}$, respectively. (e-g) Band structure with considering SOC of h-I$_{\uparrow}$, h-II$_{\uparrow}$ and h-III$_{\uparrow}$, respectively. \cite{65}}
\end{center}
\end{figure}

Then, Du et al. found that the anomalous valley Hall effect in monolayer MnPSe$_3$ can be readily reversed on or off by switching the ferroelectric polarization of monolayer Sc$_2$CO$_2$. \cite{65} As illustrated in Fig. 9(a), three typical stacking modes between monolayer MnPSe$_3$ and Sc$_2$CO$_2$, where the polarization of Sc$_2$CO$_2$ points away from the interface. The band structure including the SOC effect is illustrated in Fig. 9(b-d). For the h-I$_{\downarrow}$ and h-II$_{\downarrow}$ configuration, the feature of valley spin splitting of MnPSe$_3$ remain unchanged at the K and $\rm K'$ valleys, but the valley spin splitting values at K and $\rm K'$ valleys are no longer degeneracy. The energetic degeneracy of the K and $\rm K'$ valleys in MnPSe$_3$ is elevated, resulting in the spontaneous valley polarization. The coexistence of valley polarization and valley spin splitting ensure that anomalous valley Hall effect is observed in antiferromagnetic MnPSe$_3$ monolayer. In the h-III$_{\downarrow}$ configuration, the nonuniform potentials introduced to Mn$_1$ and Mn$_2$ atoms by the proximity effect would be rather weak, leading to the small valley spin splitting. When the ferroelectric is switched, all three configurations occur a semiconductor-to-metal transition. As shown in Fig. 9(e-g), this transition is also accompanied with the disappearance of the anomalous valley Hall effect and valley physics. Therefore, the anomalous valley Hall effect is ferroelectrically controllable, which is conducive to the development of controllable valleytronic devices.

\subsection{Interlayer stacking mode modulation}
The modulation layer stacking arrangement can realize the control of material properties, such as electronic, magnetic, ferroelectric, and superconductivity. \cite{66,67,68,69,70} Valley polarization also tune by interlayer stacking mode. \cite{71,72} For the monolayer VSi$_2$P$_4$, it exhibit ferromagnetic with the spontaneous valley polarization. Zhang et al. stack two monolayers together in the AA mode to construct a bilayer structure. \cite{73} In the bilayer structure, antiferromagnetism usually dominates the interlayer exchange interaction due to competition between the different interlayer orbital hybrids. Combined with the protection of M$_z$ symmetry, the spontaneous valley polarization is prohibited. For the resulting bilayer structure, a large Berry curvatures with opposite signs would be acquired near the K and $\rm K'$ valleys, respectively. Since the K and $\rm K'$ valleys come from the spin-down channel of the lower layer and the spin-up channel of the upper layer, respectively.

\begin{figure}[htb]
\begin{center}
\includegraphics[angle=0,width=0.9\linewidth]{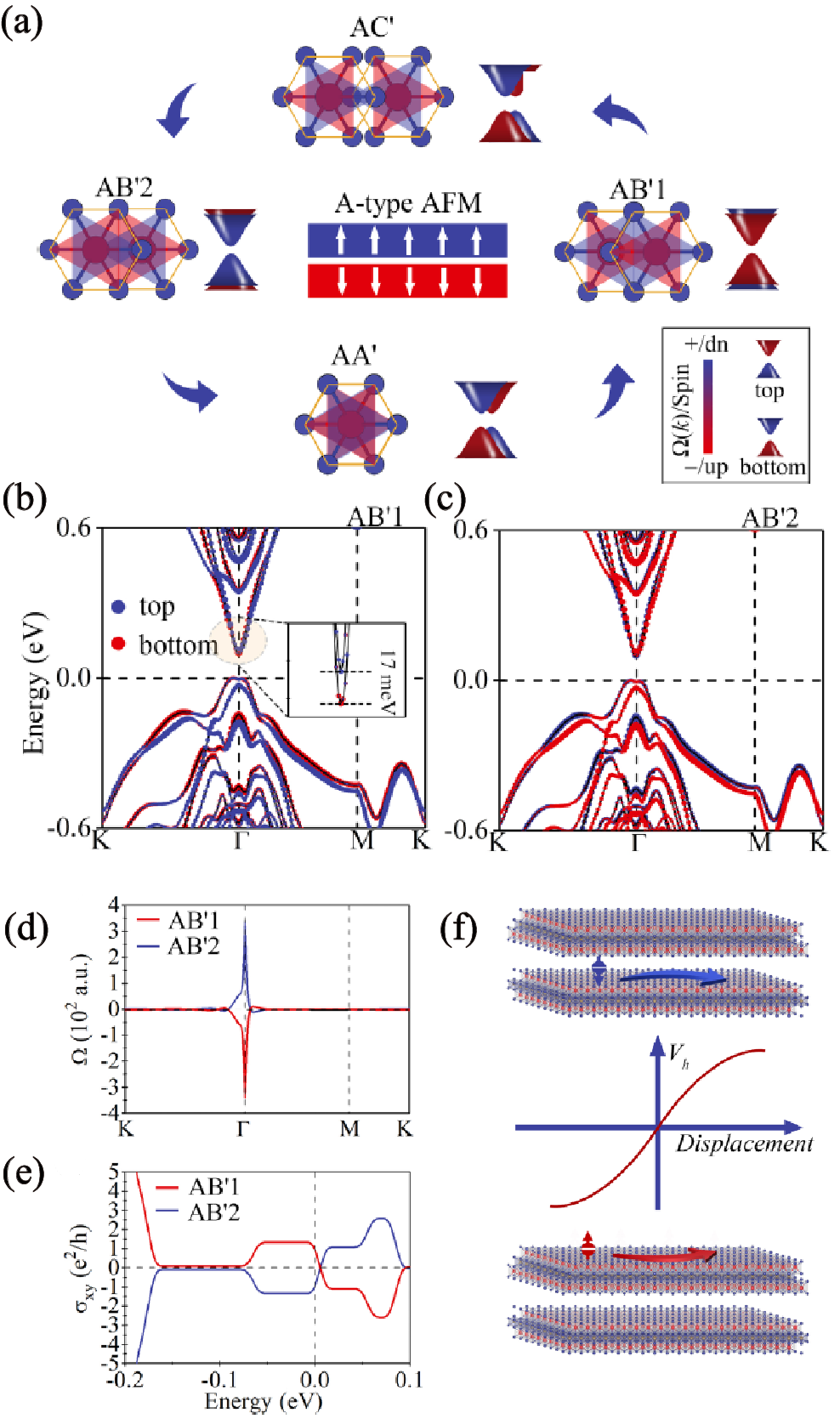}
\caption{Schematic representations of AA', AB'1, AB'2 and AC' modes of a bilayer structure and their low-energy band dispersions. Band structures for (b) AB'1 and (c) AB'2 configuration of bilayer MnBi$_2$Te$_4$ with SOC, respectively. (d) Berry curvatures along high-symmetry points for AB'1 and AB'2 configurations of bilayer MnBi$_2$Te$_4$. (e) The AHC for AB'1 and AB'2 configurations of bilayer MnBi$_2$Te$_4$. (f) Schematic diagrams of the LP-AHE for bilayer MnBi$_2$Te$_4$. \cite{72}}
\end{center}
\end{figure}

Under an interlayer translation operation of t$_1$$\|$$[-\frac{2}{3},-\frac{1}{3},0]$, t$_2$$\|$$[\frac{1}{3},-\frac{1}{3},0]$ or t$_3$$\|$$[\frac{1}{3},\frac{2}{3},0]$(-t$_1$$\|$, -t$_2$$\|$ or -t$_3$$\|$), the AA mode could be transformed into an AB (BA) mode when M$_z$ symmetry is broken. In AB and BA modes, the absence of M$_z$ symmetry would cause to an out-of-plane electrical polarization. The presence of electric polarization breaks the energetic degeneracy between the K and $\rm K'$ valleys. Therefore, the Berry curvature can be manipulated on a given layer to facilitate the realization of the layer-polarized anomalous Hall effect (LP-AHE). Besides, AB and BA modes are energetically degenerate with opposite electric polarizations, which can be viewed as two FE states. Under the action of interlayer sliding, the two FE states can be reversibly switched, resulting in an interesting sliding ferroelectric state. After the ferroelectric switching, the sign of valley polarization and electric polarization can be reversed at the same time, which has the FE control potential of layer-locked physics.

The low-energy band dispersions near the K and $\rm K'$ valleys for AB and BA modes are acquired from the k$\cdot$p model. Evidently, the spontaneous valley polarization is realized in AB and BA modes. For the AB (BA) mode, the VBM of the K ($\rm K'$) valley comes from the spin-up (spin-down) channel of the upper (lower) layer, while the CBM of the $\rm K'$ (K) valley comes from the spin-down (spin-up) channel of the lower (upper) layer. When the Fermi level shift between the K and $\rm K'$ valleys in an in-plane external electric field, the spin-down holes from the K valley can transversely move to the right edge of the upper layer, resulting in the LP-AHE in AB-stacked bilayer VSi$_2$P$_4$. Unlike the AB mode, the holes in the $\rm K'$ valley for the BA pattern would acquire a inversed anomalous velocity of -$\upsilon_a$. While the Fermi level moves between the K and K' valleys, the spin-up holes from the $\rm K'$ valley would be accumulated at the left edge of the lower layer under an in-plane electric field, forming the LP-AHE. This case is also applied to the valleys of conduction band.

Moreover, Peng et al. also found the LP-AHE in bilayer MnBi$_2$Te$_4$. \cite{72} As shown in Fig. 10(a), two configurations can be regarded as non-polar intermediate states, corresponding to two ferroelectric transition paths between AB'1 and AB'2. Fig. 10(b, c) exhibit the band structures, the bands from the bottom and top layers are separated for the AB'1 and AB'2 configurations. In addition, the calculated Berry curvature and the AHC are illustrated in Fig. 10(d, e), it can be confirm the LP-AHE in bilayer MnBi$_2$Te$_4$. By shifting the Fermi level, the electrons would be obtained at the right edge of the top layer under an in-plane electric field, as show in Fig. 10(f). The new phenomena and mechanism provide a import new way to realizing the LP-AHE and exploring its application in electronics.

\subsection{Strain modulation}
Some physical parameters of two-dimensional ferrovalley, such as valley polarization, orbit arrangement and so on, are very sensitive to lattice deformation caused by strain. It is well known that the ferrovalley property mainly originated from the broken symmetry, strain will aggravate symmetry breaking. Huan et al. found that the strain can induce half-valley metals and topological phase transitions in MBr$_2$ (M = Ru, Os) monolayer. \cite{51} As shown in Fig. 11(a-e), band structure of RuBr$_2$ can be effectively tuned between ferrovalley insulator and half-valley metal many times with the compressive strain. More interestingly, when 1.6$\% < \eta < 2.8\%$, the material turns into a Chern insulating state from ferrovalley insulator. The nature of valley related topological phase transition is the change of band gap at the K$_{+}$ and K$_{-}$ points under the compressive strain. It is summarized in Fig. 11(f). The phase transition from an ferrovalley insulator to a quantum anomalous Hall insulator and then to an ferrovalley insulator again can be obviously found.

\begin{figure}[htb]
\begin{center}
\includegraphics[angle=0,width=0.9\linewidth]{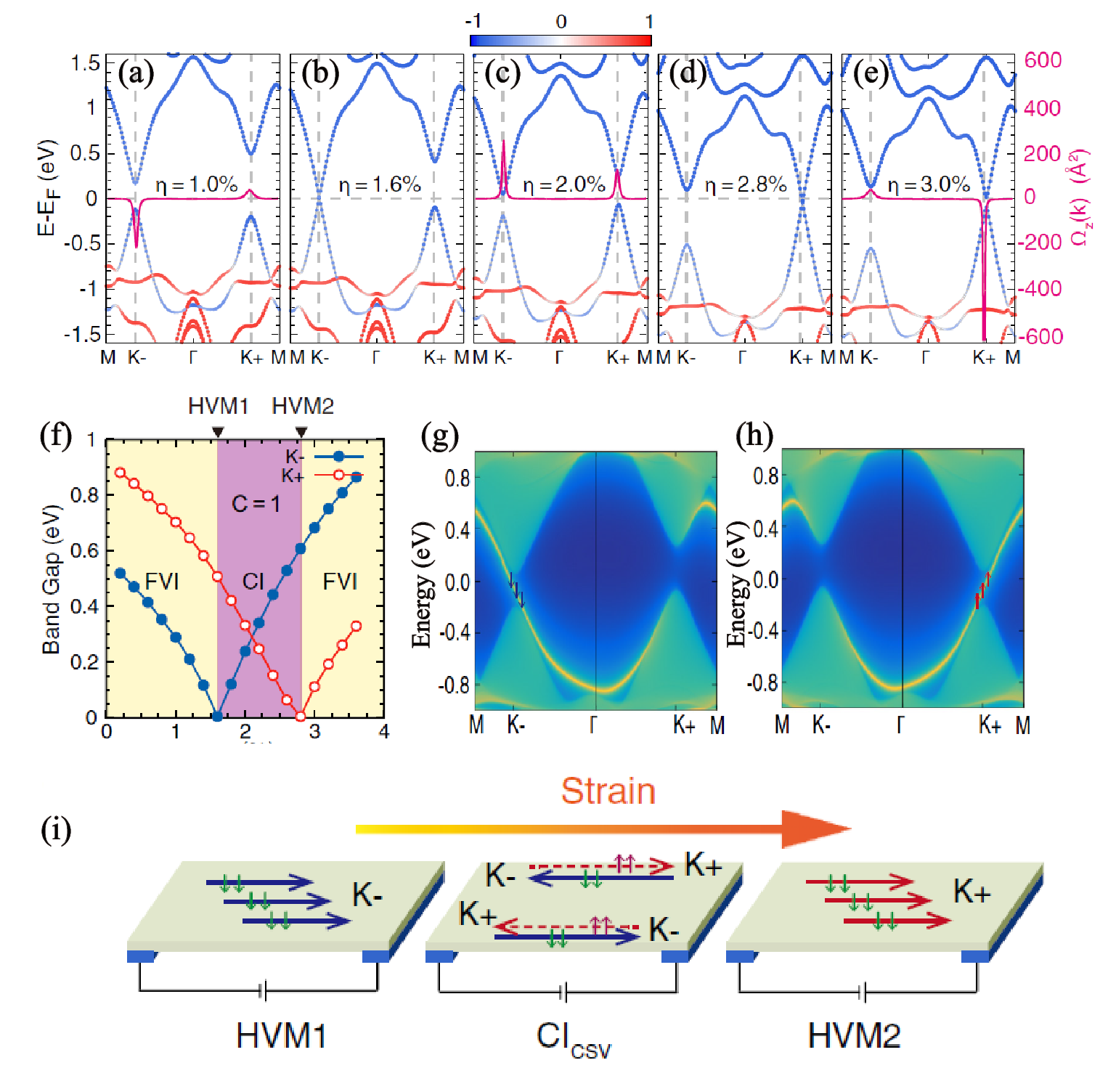}
\caption{(a-e) Band structures of the monolayer RuBr$_2$ including the SOC effect under the compressive strain. The magenta curves represent the Berry curvatures. (f) The phase diagram of the monolayer RuBr$_2$ as a function of compressive strain. (g, h) The calculated edge states for a semifinite RuBr$_2$ monolayer at the 2$\%$ compressive strain. The magnetic moment along z and -z, respectively. (i) Schematic representations of the electronic device prototypes made of the monolayer RuBr$_2$ modulated by the compressive strain. \cite{51} }
\end{center}
\end{figure}

In particular, the RuBr$_2$ monolayer becomes an half-valley metal with the compressive strain $\eta$ = 1.6 or 2.8$\%$. As shown in Fig. 11(g, h), a clear edge state connects the bulk valence and conduction bands for the 2$\%$ strain. When the magnetization is switched, the calculated edge state changes from the K$_{-}$ to K$_{+}$ valley, as an opposite chirality and spin direction. It means that the chiral-spin-valley was locked for the acquired edge state.

There rich valley and topological phase transition properties provide broad prospects for the application of electronic device. As shown in Fig. 11(i), without the strain, the system can exhibit anomalous charge/spin/valley Hall effect. While the small compressive strain is employed, 100$\%$ spin- and valley-polarization devices can be established based on the half-valley metal states acquired. On this basis, a novel Chern insulator with an ferrovalley edge state is obtained. When larger compressive strain is used, the material again transform into an half-valley metal but conducting by another valley. It shows that strain engineering can be used as an effective method to modulate the valley and topological properties of the new ferrovalley materials.

\subsection{Magnetic interface modulation}
The abundant physical phenomena was derived at the interface of two materials with different properties. The interface neighbor effect is an effective way to modulate the valley properties of ferrovalley materials. Ma et al. constructed the 2H-VS$_2$/Cr$_2$C heterostructures, as shown in Fig. 12(a-d). \cite{74} They investigated the effects of four different magnetic configurations on valley polarization. In the four configurations, all the heterojunctions are semiconducting. In Fig. 12(a, b), the VBM band is the spin-down channel. As shown in Fig. 12(c,d), while the magnetization is switched, the VBM band change to the spin-up channel. Moreover, while the direction of magnetization is out-of-plane, as illustrated in Fig. 12(a, c), the valley split has the opposite sign but the same value because of the opposite direction of magnetization. More interestingly, while the direction of magnetization turns into in-plane, as shown in Fig. 12(b ,d), the valley polarization of K and $\rm K'$ points disappears. In brief, the switch direction of magnetization causes the valley polarization and spin in reverse.

\begin{figure}[htb]
\begin{center}
\includegraphics[angle=0,width=0.9\linewidth]{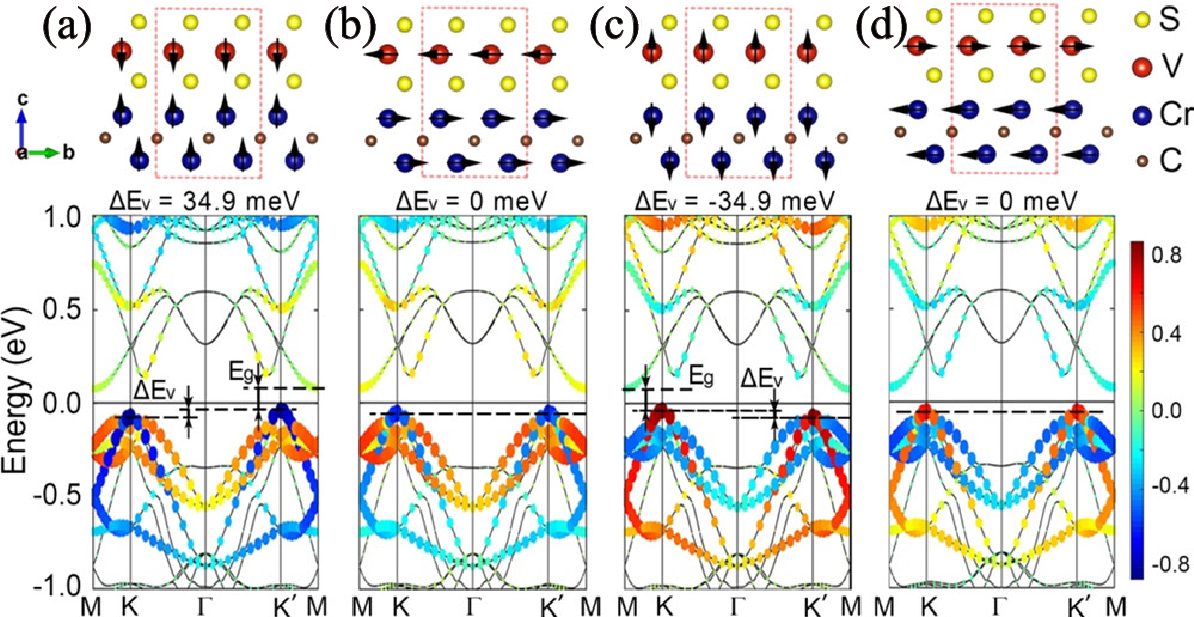}
\caption{(a-d) Four different magnetization configurations of 2H-VS$_2$/Cr$_2$C and its energy structure. The arrows represent the direction of magnetic moments, and the red dotted line box exhibits the atomic structure of heterjunctions. The circles present the e$_2$ (d$_{xy}$ + d$_{x2-y2}$) orbital of VS$_2$, the color scale represent the spin projection. \cite{74} }
\end{center}
\end{figure}

\section{Summary and Outlook}
In summary, we introduce the investigation progress of two-dimensional ferrovalley materials. Firstly, we introduce the background of research related to valley. Then, the basic concepts of valleytronics, valley freedom and valley polarization are introduced through the spatial inversion of broken graphene and 2H-TMDs materials. Following, we mainly discuss the ferrovalley materials formed by different orbits. Finally, the multi-field tunable of two-dimensional ferrovalley materials is concerned.

Even though two-dimensional ferrovalley materials have been predicted early by the DFT calculation, the experimental progress has been limited, partly related to the lack of powerful sample characterization and preparation technique. Recently, Guan et al. firstly reported the successful preparation of intrinsic bulk ferrovalley material in experiment. \cite{75} For two-dimensional ferrovalley, so far, there has been no experimental progress.

The key of electronic device design is to construct binary logic switching state. Similar to spin, valley is a new type of freedom, different valley polarization states can be used to encode and store information. At present, spintronics devices are mainly based on ferromagnetic materials. The investigation of electronic devices based on spin injection semiconductor materials is still in its infancy. Ferrovalley has strong locking between valley and spin, it provides a new platform for the study of spintronics. Moreover, a large number of valley-based spintronic devices have been explored. For example, using valley to transmit information has high fidelity, direction and room temperature operability. \cite{76} More importantly, the neural network calculation using valley transistor has high accuracy. \cite{77} The spin based on valley is more stable, and the spin control and detection can be realized with the help of valley, which expands the application prospect of the device.

\section*{ACKNOWLEDGEMENTS}
This work is supported by the National Natural Science Foundation of China (Grant No. 12074301, No. 12004295). P. Li thanks China's Postdoctoral Science Foundation funded project (Grant No. 2022M722547), and the Open Project of State Key Laboratory of Surface Physics (No. KF2022$\_$09). W. Zhang thanks the Natural Science Foundation of Guizhou Provincial Provincial Education Department (Grant No. ZK[2021]034).


\end{document}